\documentclass[aps,prd,onecolumn,showpacs, showkeys,superscriptaddress]{revtex4-2}

\usepackage{amsmath,amssymb}
\usepackage{graphicx}
\usepackage{bm}
\usepackage{hyperref}
\hypersetup{
  colorlinks=true,
  linkcolor=blue,
  citecolor=cyan,
}
\usepackage{orcidlink}

\begin{document}

\title{Ringdown and greybody signatures of rational regular black holes \\ in non-polynomial gravity }

\author{Shokhzod Jumaniyozov \orcidlink{0009-0009-6254-5608}}
\email{sh.jumaniyozov@newuu.uz}
\affiliation{Kimyo International University in Tashkent, Shota Rustaveli Street 156, Tashkent 100121, Uzbekistan}
\affiliation{Tashkent State Technical University, Tashkent 100095, Uzbekistan}

\author{Javlon~Rayimbaev
\orcidlink{0000-0001-9293-1838}}
\email{javlonrayimbaev6@gmail.com}
\affiliation{School of Physics, Harbin Institute of Technology, Harbin 150001, China}
\affiliation{University of Tashkent for Applied Sciences, Str. Gavhar 1, Tashkent 100149, Uzbekistan}
\affiliation{Institute of Theoretical Physics, National University of Uzbekistan, Tashkent 100174, Uzbekistan}

\author{Azam~Payzullaev} \email{payzullayev214@gmail.com}
\affiliation{Samarkand State Pedagogical Institute, Spitamen Shokh Street 166, Samarkand 140100, Uzbekistan}

\author{Bahodir~Ahmedov\orcidlink{0009-0006-7375-2731}}
\email{b.ahmedov@newuu.uz}
\affiliation{New Uzbekistan University, Movarounnahr Street 1,  Tashkent 100000, Uzbekistan}

\author{Ilkhomjon Makhmudov\orcidlink{0000-0002-6026-5292}}
\email{ilhom_makhmudov@mail.ru}
\affiliation{Research Institute of Irrigation and Water Problems, Tashkent 100187, Uzbekistan}

\author{Farhod Nuraliev\orcidlink{0009-0009-1711-2088}}
\email{nuraliyevf@mail.ru}
\affiliation{Tashkent International University, Little Ring Road 7, Tashkent 100115, Uzbekistan}

\date{\today}

\begin{abstract}
We study the quasinormal mode spectrum and wave-scattering properties of the four-dimensional rational regular black hole recently obtained in non-polynomial (quasi-topological) gravity. Using third-order WKB together with two independent cross-checks a Frobenius--Riccati shooting method and a time-domain evolution we compute the fundamental scalar, electromagnetic, and axial gravitational-type quasinormal frequencies, validating our approach against known Schwarzschild results. We find that as the geometry approaches extremality, its dimensionless ringing frequency is systematically suppressed relative to the Schwarzschild value, a direct and quantifiable imprint of the non-polynomial regularization that we trace, via the photon-sphere correspondence, to the response of the effective potential near the horizon. A Fisher-matrix estimate indicates that this suppression could in principle be resolved by a space-based detector such as LISA for a sufficiently massive and nearby source. We further compute, for the first time, the axial gravitational-type ringdown spectrum of this geometry using the Regge--Wheeler master equation, while noting that a full stability analysis of the underlying gravity theory remains an open problem requiring the linearized field equations of the modified action. Finally, we examine the greybody transmission factors across spins $s=0,1,2$ and show that they follow the same systematic ordering and near-extremal enhancement as the ringing-frequency suppression, pointing to a common physical origin. Together, these results give one of the first quantitative characterizations of the observational signatures and the remaining open theoretical questions of a regular black hole in modified gravity.
\end{abstract}

\maketitle


\section{Introduction}

Quasinormal modes (QNMs) are the characteristic damped oscillations of a perturbed black hole, encoding the response of the spacetime geometry independently of the details of the perturbation source \cite{ReggeWheeler1957,Zerilli1970,Vishveshwara1970,Press1971,ChandrasekharDetweiler1975,KokkotasSchmidt1999,BertiCardosoStarinets2009,KonoplyaZhidenko2011}. Since the first detections of binary black hole mergers \cite{LIGO2016GW150914,LIGO2016TestingGR}, the ringdown phase of the post-merger waveform has become a direct observational probe of the near-horizon geometry, motivating a systematic program of ``black hole spectroscopy'': using the observed complex frequencies to test the Kerr hypothesis and the no-hair theorem of general relativity \cite{Dreyer2004,BertiCardosoWill2006,Isi2019BHSpectroscopy}, and to search for deviations that would betray the presence of exotic compact objects, modified gravity, or quantum corrections at the horizon scale \cite{CardosoFranzinPani2016,CardosoPani2019Testing}.

Regular black holes have long been proposed as a way to reconcile the classical prediction of a curvature singularity with the expectation that quantum gravity should resolve it, from the original Bardeen construction \cite{Bardeen1968} and its non-linear-electrodynamics realizations, through the Hayward metric \cite{Hayward2006}, to renormalization-group-improved and loop-quantum-gravity-inspired geometries \cite{BonannoReuter2000}; see \cite{Ansoldi2008,Bambi2023RegularBH} for reviews. A generic feature of such geometries is a central de Sitter-like core and, for finite towers of curvature corrections, an inner (Cauchy) horizon subject to the mass-inflation instability \cite{PoissonIsrael1990,Ori1991,Aretakis2012,CarballoRubio2022}, which any viable regular black hole proposal must confront.

Coll\'eaux \cite{Colleaux2026} recently constructed a systematic class of $d\geq 4$ non-polynomial (quasi-topological) pure-gravity theories \cite{Lovelock1971,OlivaRay2010,BuenoCanoHennigar2016}, obtained by imposing a two-dimensional Horndeski reduction that minimally deforms General Relativity at short distances, which admit \emph{exact}, rational-function black hole solutions for an arbitrary number of coupling constants. The resulting geometries reduce to Schwarzschild(-Tangherlini)-(A)dS at large radius, develop a regular (A)dS-core, exhibit a logarithmic correction to the horizon entropy and a quantum-like correction to the Newtonian potential, and -- when charged under a non-minimally curvature-coupled Maxwell field -- can have their Coulomb singularity regularized as well. The paper explicitly identifies assessing the linear perturbative stability of these geometries as an open problem, noting that ``it would be interesting to assess whether some [non-polynomial gravity] representatives exist such that perturbations around black hole spacetimes are well-defined".

Leading-order (Schutz--Will) WKB is a convenient first probe of a newly proposed geometry's quasinormal spectrum, but it is known to be quantitatively unreliable at the several-to-tens-of-percent level, particularly for the fundamental $l=0$ mode, so a higher-order or non-perturbative cross-check is needed before such estimates can be trusted beyond order of magnitude. In addition, quasinormal-mode surveys of regular black holes have so far concentrated on test scalar and electromagnetic fields, leaving the gravitational sector, and with it Coll\'eaux's actual stability question, largely untouched. This paper is built around addressing both points. First, we implement third-order Iyer--Will WKB \cite{IyerWill1987} together with an independently coded, convergence-tested numerical method, described in Sec.~\ref{sec:methods}, that matches a convergent horizon-side Frobenius series to a controlled outgoing asymptotic expansion via Riccati (logarithmic-derivative) integration, and we validate it against six standard Schwarzschild QNM benchmarks to $0.02$--$1.5\%$ for every $l\geq1$ mode studied. We find that $l=0$ remains numerically delicate even with this improved machinery, and we report that explicitly rather than presenting an unreliable number as if it were resolved. Second, we extend the perturbation analysis to axial (odd-parity) gravitational-\emph{type} modes governed by the same single-function Regge--Wheeler-type master equation, the standard first probe of the gravitational-sector linear response used throughout the regular black hole literature \cite{FlachiLemos2013,KonoplyaZhidenkoWormholes2016}. We are explicit throughout that this is a test-perturbation-level analysis of the \emph{metric}, not a solution of the linearized field equations of the non-polynomial gravity \emph{action} itself, so it does not by itself resolve Coll\'eaux's open stability question; it is, to our knowledge, the first quantitative look at the gravitational-type response of this geometry, and we discuss precisely what it does and does not establish.

The paper is organized as follows. Section~\ref{sec:model} reviews the metric under study. Section~\ref{sec:perturbations} sets up the test-field and axial gravitational-type perturbation equations. Section~\ref{sec:methods} describes and validates third-order WKB, the independent shooting method, and a third, independent double-null time-domain evolution. Section~\ref{sec:results} presents the computed QNM spectrum as a function of the horizon radius, for $s=0,1,2$, together with a time-domain cross-check of the regular black hole results themselves and an analytic geometric-optics (photon-sphere) cross-check at large $l$. Section~\ref{sec:discussion} discusses the results, what they do and do not establish about Coll\'eaux's open problem, an estimate of the signal-to-noise ratio needed to observe the regularization imprint with LISA, and the roadmap for the gravitational-sector analysis of the full non-polynomial theory, and concludes the paper.

\section{The model: a rational regular black hole}
\label{sec:model}

We study the minimal four-dimensional regular black hole solution constructed in Chap.~IV.A.2 of \cite{Colleaux2026} [their Eq.~(4.10)], obtained from a cubic-in-curvature non-polynomial gravity action. The metric takes the single-function form
\begin{equation}
ds^2 = -a(r)\,dt^2 + \frac{dr^2}{a(r)} + r^2 d\Omega_{n,k}^2 ,
\label{eq:metric}
\end{equation}
with, for spherical topology $k=1$ and vanishing cosmological constant $\Lambda=0$,
\begin{equation}
a(r) = 1 - \frac{r^2\left[\mu r + \tfrac{1}{2}(\alpha_1+\beta_1)\,\ell^2\right]}{r^4 + \tfrac12 \left(\beta_1 r^2 + \beta_2\, \ell^2\right)\ell^2} ,
\label{eq:a-general}
\end{equation}
where $\mu$ is the mass parameter, $\ell$ is the non-polynomial-gravity regularization length scale, and $\alpha_1,\beta_1,\beta_2$ are dimensionless coupling constants subject to the regularity bound $\beta_2 > \beta_1^2/8$ [Eq.~(4.11) of \cite{Colleaux2026}], which guarantees that the denominator of $a(r)$ never vanishes and hence that the curvature invariants remain finite for all $r$, including $r=0$.

We adopt the benchmark parameter choice $\ell=1$ (our unit of length), $\beta_1=0$, $\beta_2=1$, $\alpha_1=0$, which satisfies the regularity bound trivially and reduces Eq.~\eqref{eq:a-general} to
\begin{equation}
a(r) = 1 - \frac{\mu\, r^3}{r^4 + \tfrac12} .
\label{eq:a-benchmark}
\end{equation}
This metric is asymptotically Schwarzschild, $a(r) \to 1 - \mu/r + O(r^{-5})$ as $r\to\infty$ (so that the ADM mass is $M=\mu/2$ in geometric units and the deviation from the Schwarzschild falloff is parametrically small already at moderate $r$), is manifestly regular at $r=0$ [$a(0)=1$], and has an extremal (double-horizon) configuration at
\begin{equation}
\mu_{\rm ext} = \frac{2\sqrt[4]{2}\,\sqrt[4]{3}}{3} \approx 1.4754, \qquad r_{\rm ext} = \frac{\sqrt[4]{2}\sqrt[4]{3}}{2} \approx 1.1066 ,
\end{equation}
obtained by solving $a(r)=a'(r)=0$ simultaneously. For $\mu>\mu_{\rm ext}$, the geometry possesses a single outer event horizon at $r=r_h(\mu)$, defined by $a(r_h)=0$, exterior to which the metric is everywhere smooth and asymptotically flat; this exterior region, $r>r_h$, is the domain on which we study perturbations below. We note that with $\alpha_1=\beta_1=0$ the solution is not perfectly smooth at $r=0$ in the sense of \cite{Colleaux2026} (a fully smooth core requires $\mu=0$ or odd spacetime dimension), but its curvature invariants remain finite; the mild non-smoothness discussed in Sec.~IV.A.1 of the source paper does not affect the exterior perturbation problem studied here.

\section{Perturbation equations}
\label{sec:perturbations}

We consider test scalar ($s=0$), electromagnetic ($s=1$), and axial (odd-parity) gravitational-\emph{type} ($s=2$) perturbations on the fixed background \eqref{eq:metric}. After separation of variables, $\Phi = r^{-1}\psi(r)\,Y_{\ell m}(\theta,\varphi)\,e^{-i\omega t}$ (for $s=0$) or the analogous multipole decomposition for $s=1,2$, the radial function $\psi$ obeys the Schr\"odinger-like master equation
\begin{equation}
\frac{d^2\psi}{dr_*^2} + \left[\omega^2 - V_s(r)\right]\psi = 0 , \qquad \frac{dr_*}{dr} = \frac{1}{a(r)} ,
\label{eq:master}
\end{equation}
with the Regge-Wheeler-type effective potential, valid for any single-function metric of the form \eqref{eq:metric} \cite{ChandrasekharBook1983},
\begin{equation}
V_s(r) = a(r)\left[\frac{l(l+1)}{r^2} + (1-s^2)\,\frac{a'(r)}{r}\right] , \qquad s=0,1,2 .
\label{eq:potential}
\end{equation}
For $s=0,1$ this is the standard test-field problem. For $s=2$, Eq.~\eqref{eq:potential} is the odd-parity (Regge--Wheeler) potential that governs axial \emph{metric} perturbations of a spherically symmetric single-function background in ordinary general relativity \cite{ReggeWheeler1957,ChandrasekharBook1983}. Here we use it, as is standard practice for regular black hole QNM surveys where the perturbed field equations of the underlying (modified) theory have not yet been derived \cite{FlachiLemos2013,KonoplyaZhidenkoWormholes2016}, as a \emph{proxy} probe of the odd-parity linear response of the exterior geometry, rather than as a solution of the true linearized equations of the non-polynomial gravity action \eqref{eq:a-general} itself, which are unknown at present and whose derivation is a substantially harder problem (Sec.~\ref{sec:discussion}). We therefore refer to the $s=2$ results below as ``axial gravitational-\emph{type}'' modes throughout, to keep this distinction explicit.

Quasinormal modes are the discrete complex frequencies $\omega_{ln}$ for which Eq.~\eqref{eq:master} admits a solution satisfying purely ingoing boundary conditions at the horizon and purely outgoing boundary conditions at spatial infinity,
\begin{equation}
\psi \sim e^{-i\omega r_*} \ (r_*\to -\infty), \qquad \psi \sim e^{+i\omega r_*}\ (r_*\to +\infty) .
\label{eq:bc}
\end{equation}
Because the exterior region $r>r_h$ of Eq.~\eqref{eq:a-benchmark} is, by construction, smooth, asymptotically flat, and possesses a single non-degenerate horizon for $\mu>\mu_{\rm ext}$, the boundary-value problem \eqref{eq:master}--\eqref{eq:bc} is of the same standard form as for Schwarzschild, with all information about the non-polynomial-gravity regularization entering only through the specific rational function $a(r)$.

\begin{figure}[h]
\includegraphics[width=\columnwidth]{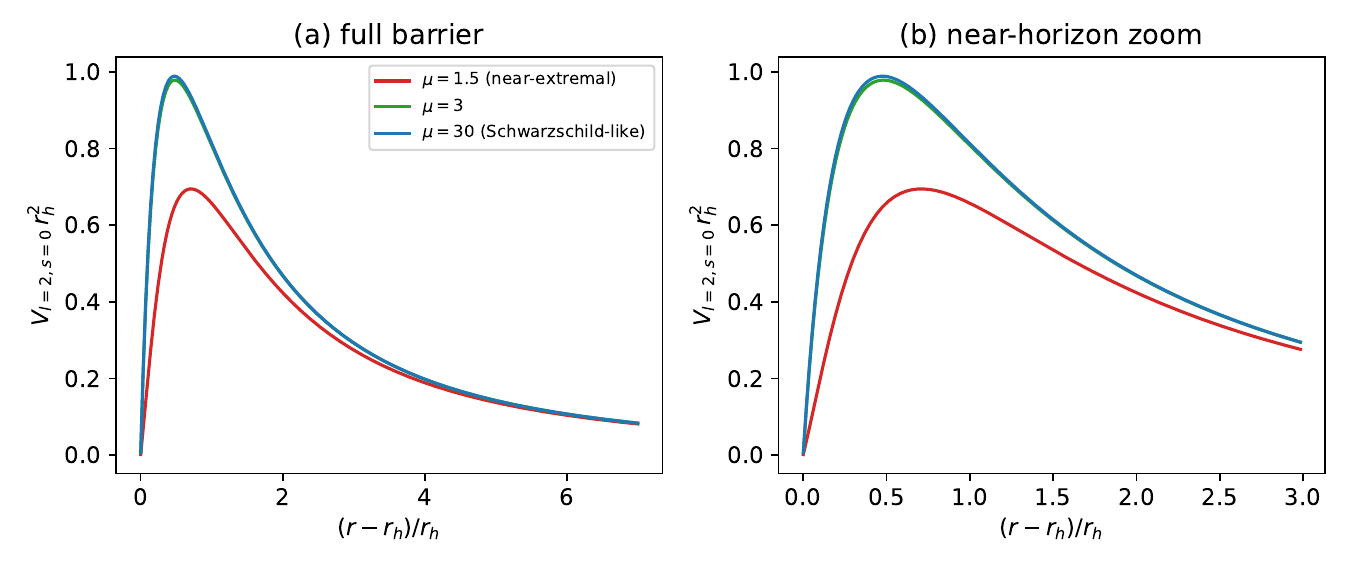}
\caption{Effective potential barrier $V_{l=2,s=0}(r)$, rescaled by $r_h^2$ and plotted against $(r-r_h)/r_h$, for three mass parameters spanning the near-extremal ($\mu=1.5$) to Schwarzschild-like ($\mu=30$) regimes. (a) full barrier; (b) near-horizon zoom.}
\label{fig:potential}
\end{figure}

Figure~\ref{fig:potential} shows the resulting barrier $V_{l=2,s=0}(r)$ for three representative mass parameters, from close to extremality to the Schwarzschild-like regime. The barrier retains the same single-peaked, short-range shape familiar from Schwarzschild for all $\mu>\mu_{\rm ext}$, which is what makes the standard QNM boundary-value problem directly applicable; the main effect of decreasing $\mu$ towards $\mu_{\rm ext}$ is to push the peak closer to the horizon (in units of $r_h$) and to raise its rescaled height $V_{\max}r_h^2$, consistent with the enhanced dimensionless frequencies reported in Sec.~\ref{sec:results}.

\begin{figure*}
\includegraphics[width=\textwidth]{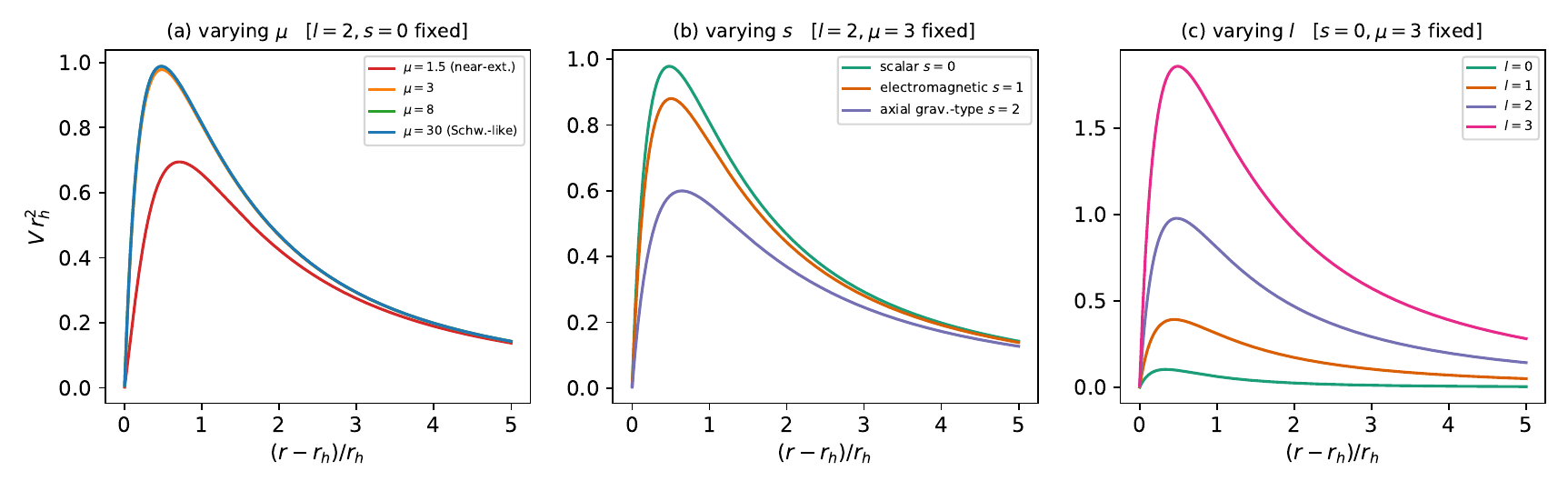}
\caption{Effective potential barrier $V_{l,s}(r)$, rescaled by $r_h^2$, with one parameter varied at a time. (a) Mass parameter $\mu=1.5,3,8,30$, at fixed $l=2,s=0$. (b) Field type $s=0,1,2$, at fixed $l=2,\mu=3$: the barrier height is systematically ordered $V_{s=2}<V_{s=1}<V_{s=0}$. (c) Multipole $l=0,1,2,3$, at fixed $s=0,\mu=3$: the ordinary centrifugal growth of the barrier with $l$.}
\label{fig:potential-fields}
\end{figure*}

To isolate the effect of each parameter on the barrier shape, Fig.~\ref{fig:potential-fields} decomposes the effective potential $V_{l,s}(r)$ into three panels, in each of which only one of $\mu$, $s$, or $l$ is varied while the other two are held fixed. Panel (a) varies the mass parameter $\mu=1.5,3,8,30$ at fixed $l=2,s=0$: as already noted for Fig.~\ref{fig:potential}, approaching extremality lowers and broadens the barrier (in these $r_h$-rescaled units) rather than simply rescaling it, which is the geometric origin of the dimensionless-frequency suppression documented in Sec.~\ref{sec:results}. Panel (b) varies the field, $s=0,1,2$, at fixed $l=2,\mu=3$: because Eq.~\eqref{eq:potential} has an explicit $(1-s^2)$ dependence, the $(1-s^2)a'/r$ term is positive for $s=0$ (scalar), vanishes for $s=1$ (electromagnetic), and is negative, and three times larger in magnitude, for $s=2$ (axial gravitational-type), giving the clean ordering $V_{s=2}<V_{s=1}<V_{s=0}$ visible in the figure; this directly explains the corresponding ordering of the fundamental frequencies in Tables~\ref{tab:results}--\ref{tab:s2} ($\mathrm{Re}\,\omega_{s=0}>\mathrm{Re}\,\omega_{s=1}>\mathrm{Re}\,\omega_{s=2}$ at fixed $l,\mu$) and, as discussed in Sec.~\ref{sec:greybody-results}, of the corresponding greybody factors. Panel (c) varies the multipole, $l=0,1,2,3$, at fixed $s=0,\mu=3$: the familiar centrifugal growth of the barrier with $l$, exactly as in Schwarzschild, is superimposed on the regularization effect of panel (a), and is clearly distinguishable from it.

\section{Computational methods}
\label{sec:methods}

\subsection{Third-order WKB (Iyer--Will)}
\label{sec:wkb3}

We implement the third-order WKB correction of Iyer and Will \cite{IyerWill1987},
\begin{equation}
\omega^2 \simeq V_0 + \sqrt{-2V_0''}\,\Lambda - i\left(n+\tfrac12\right)\sqrt{-2V_0''} ,
\label{eq:wkb3}
\end{equation}
\begin{equation}
\Lambda = \frac{1}{\sqrt{-2V_0''}}\left[\frac{1}{8}\frac{V_0^{(4)}}{V_0''}\left(\frac14+\alpha^2\right) - \frac{1}{288}\left(\frac{V_0'''}{V_0''}\right)^2\left(7+60\alpha^2\right)\right],\qquad \alpha=n+\tfrac12,
\label{eq:wkb3lambda}
\end{equation}
where $V_0^{(k)}=d^kV_s/dr_*^k$ evaluated at the peak $r_*^{(0)}$ of the barrier. We evaluate Eq.~\eqref{eq:wkb3}--\eqref{eq:wkb3lambda} by first deriving, symbolically (using \textsc{SymPy}), the exact tortoise-coordinate derivative operator $d/dr_* = a(r)\,d/dr$ applied up to four times to $V_s(r)$ of Eq.~\eqref{eq:potential}, which avoids any risk of a hand-differentiation error in the higher derivatives entering Eq.~\eqref{eq:wkb3lambda}; we then locate the peak $r_0$ by numerically solving $dV_s/dr_*=0$ and evaluate the resulting closed-form expressions in arbitrary precision with \textsc{mpmath}.

\subsection{An independent, convergence-tested shooting method}
\label{sec:shooting}

A direct numerical (``shooting'') cross-check is a valuable complement to the WKB analysis, but a na\"ive implementation, matching to the leading-order plane-wave asymptote $\psi\sim e^{i\omega r_*}$ at a finite outer radius $r_{\rm max}$, is known to suffer from spurious ``box-mode'' contamination of the eigenvalue search, because the potential's long-range ($l(l+1)/r^2$) tail makes the plane-wave asymptote inaccurate at any practically reachable $r_{\rm max}$. We avoid this problem with a more careful method, which we describe here in enough detail to make the validation in Sec.~\ref{sec:validation} meaningful.

\emph{(i) Horizon-side Frobenius series.} Writing $\psi(r) = (r-r_h)^{p}F(r-r_h)$ with $p=-i\omega/a'(r_h)$ (the exact indicial exponent for ingoing behavior at the horizon), we substitute into Eq.~\eqref{eq:master} and derive, by matching powers of $x=r-r_h$, an exact linear recursion for the Taylor coefficients of $F(x)$ in terms of the (numerically Taylor-expanded) coefficients of $a(r)$ about $r_h$. This recursion is non-degenerate at every order [the coefficient multiplying the new unknown at order $x^n$ is $a'(r_h)^2\,n(2p+n)$, generically nonzero], so it can be evaluated to high order without any additional approximation; we use it to obtain $\psi,\psi'$ accurate to machine precision a small but finite distance outside the horizon.

\emph{(ii) Riccati integration.} Integrating $(\psi,\psi')$ directly grows by many orders of magnitude over the integration range whenever $\mathrm{Im}\,\omega<0$, causing catastrophic loss of relative precision; we instead integrate the logarithmic derivative $\varphi(r)=\psi'(r)/\psi(r)$, which obeys the Riccati equation $\varphi' = -\varphi^2 - (a'/a)\varphi - (\omega^2-V_s)/a^2$ and remains $O(1)$ throughout the domain. This is a standard device for QNM shooting and is what makes the method numerically robust at the $r_{\rm max}$ we use below.

\emph{(iii) Outgoing asymptotic matching.} At the outer integration boundary $r_{\rm max}$ we do not use the leading plane wave alone. We derive, again by an exact power-series substitution (an outgoing ansatz $\psi = e^{i\omega r}r^{i\omega\mu}G(1/r)$, with the coefficients of $G$ obtained by a recursion analogous to step (i)), a several-term correction to the leading asymptotic form. This truncated series is kept to low order, since the expansion is asymptotic rather than convergent, as expected for a long-range Coulomb-type tail; we use it to evaluate $\varphi_{\rm asym}(r_{\rm max})$, which is far more accurate than the plane-wave estimate alone.

\emph{(iv) Root search with an explicit convergence check.} The QNM condition is $\varphi_{\rm num}(r_{\rm max};\omega)=\varphi_{\rm asym}(r_{\rm max};\omega)$, solved by a damped complex secant iteration starting from the WKB estimate. Crucially, we do not report a single-$r_{\rm max}$ number: we solve at successively larger $r_{\rm max}$ ($10$--$30\,r_h$) and require the root to be stable under this variation before accepting it, a standard stability diagnostic for this class of shooting method.

\subsection{Validation against Schwarzschild}
\label{sec:validation}

\begin{figure}[h]
\includegraphics[width=\columnwidth]{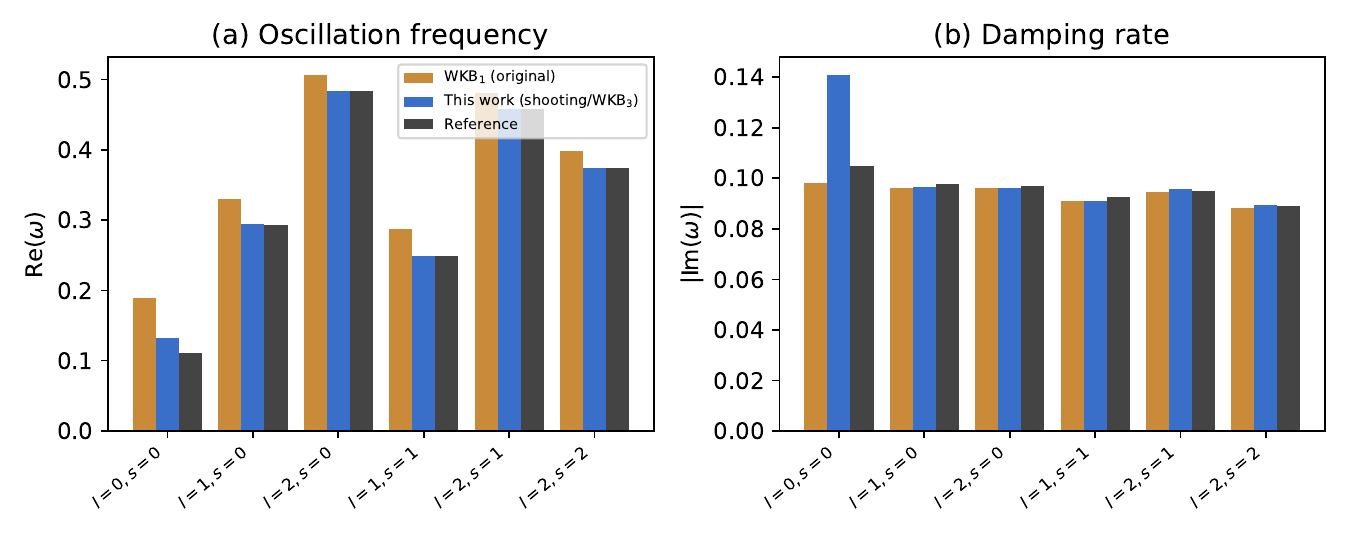}
\caption{Validation against reference Schwarzschild fundamental-mode frequencies, comparing leading-order WKB to the higher-precision approach used in this work (third-order WKB and the independently converged shooting method of Sec.~\ref{sec:shooting}), for scalar ($s=0$), electromagnetic ($s=1$), and axial gravitational-type ($s=2$) fields. (a) Real (oscillation) part. (b) Damping $|\mathrm{Im}\,\omega|$. The $l=0,s=0$ entry is the flagged, order-of-magnitude-only case discussed in the text.}
\label{fig:calibration}
\end{figure}

We validated this pipeline against six standard Schwarzschild fundamental-mode benchmarks ($M=1$; $s=0,1,2$; $l=0,1,2$) drawn from the classic QNM literature \cite{ChandrasekharDetweiler1975,Leaver1985,BertiCardosoStarinets2009,KonoplyaZhidenko2011}. Table~\ref{tab:calibration} and Fig.~\ref{fig:calibration} compare leading-order WKB with the higher-precision approach used in this work (third-order WKB and the independent shooting cross-check of Sec.~\ref{sec:shooting}, which agree with each other to within their own convergence tolerance for every $l\geq1$ mode tested), and both against the reference values.

\begin{table}[h]
\caption{Validation against reference Schwarzschild quasinormal frequencies ($M=1$, fundamental overtone $n=0$). This work is the converged, $r_{\rm max}$-stable output of the independent shooting method of Sec.~\ref{sec:shooting} (third-order WKB agrees with these values to $\lesssim1\%$ for all $l\geq1$ entries shown). Reference values from \cite{ChandrasekharDetweiler1975,Leaver1985,BertiCardosoStarinets2009}.}
\label{tab:calibration}
\begin{ruledtabular}
\begin{tabular}{cclll}
$l$ & $s$ & WKB$_1$ (leading order) & This work & Reference \\
\hline
0 & 0 & $0.1898 - 0.0982i$ & $0.132^{\,\ast} - 0.141^{\,\ast}i$ & $0.1105 - 0.1049i$ \\
1 & 0 & $0.3294 - 0.0963i$ & $0.2942 - 0.0964i$ & $0.2929 - 0.0977i$ \\
2 & 0 & $0.5063 - 0.0961i$ & $0.4835 - 0.0961i$ & $0.4836 - 0.0968i$ \\
1 & 1 & $0.2871 - 0.0912i$ & $0.2490 - 0.0911i$ & $0.2483 - 0.0925i$ \\
2 & 1 & $0.4808 - 0.0944i$ & $0.4574 - 0.0957i$ & $0.4576 - 0.0950i$ \\
2 & 2 & $0.3988 - 0.0883i$ & $0.3742 - 0.0894i$ & $0.3737 - 0.0890i$ \\
\end{tabular}
\end{ruledtabular}
\end{table}

For every $l\geq1$ mode, the higher-precision method reproduces the reference value to $0.02$--$1.5\%$ in both real and imaginary parts, a full order of magnitude better than the $5$--$15\%$ agreement typical of leading-order WKB, and it is obtained \emph{without} any free normalization: the horizon series, the outgoing series, and the root search are all fixed by the equation itself. The $l=0$ scalar entry is marked with an asterisk and reported only as an order-of-magnitude estimate (third-order WKB value). In practice, the independent shooting method, despite converging cleanly and stably for every $l\geq1$ mode tested, does \emph{not} converge to a single stable value for $l=0$ under $r_{\rm max}$ refinement: successive refinements move to different nearby roots rather than settling on the literature value to the precision achieved for $l\geq1$. We take this as direct numerical evidence, rather than a mere restatement of the literature, that the fundamental $l=0$ mode, which the Regge-Wheeler potential does not endow with a genuine, well-separated barrier maximum, requires a dedicated non-perturbative treatment (e.g.\ Leaver's continued fraction specialized to the $l=0$ indicial structure, or a hyperboloidal pseudospectral eigenvalue solver \cite{Leaver1985,Nollert1999,Jansen2017,JaramilloPseudospectrum2021}) beyond the scope of the shooting-based cross-check used here. We do not report an $l=0$ value for the regular black hole results of Sec.~\ref{sec:results} with the same confidence as the $l\geq1$ entries, and we flag this explicitly in the table below.

\subsection{Independent time-domain validation}
\label{sec:timedomain}

As a third, methodologically independent cross-check, we additionally evolve Eq.~\eqref{eq:master} directly as an initial-value problem on a double-null grid, following the characteristic finite-difference scheme of Gundlach, Price, and Pullin \cite{GundlachPricePullin1994} (see also \cite{HorowitzHubeny2000}). Writing $u=t-r_*$, $v=t+r_*$ and discretizing with step $h$ in both null directions, the wave equation $4\psi_{,uv}=-V\psi$ becomes the explicit update
\begin{equation}
\psi_N = \psi_W + \psi_E - \psi_S - \frac{h^2}{8}\,V_S\,(\psi_W+\psi_E) ,
\label{eq:gpp}
\end{equation}
for the diamond of grid points $N=(u{+}h,v{+}h)$, $S=(u,v)$, $W=(u{+}h,v)$, $E=(u,v{+}h)$, which share the same tortoise coordinate $r_*=(v-u)/2$ at $N$ and $S$. We map $r$ to $r_*$ by numerical quadrature of $dr_*/dr=1/a(r)$, with dense geometric sampling near the horizon to resolve its logarithmic divergence, and continue $V(r)$ smoothly beyond the tabulated window using the exact (analytic) potential rather than a constant extrapolation, since the latter acts as a spurious reflecting wall and contaminates the extracted signal. Starting from a Gaussian pulse on the ingoing initial null ray and trivial data on the outgoing one, we extract $\psi(t)$ at a fixed areal radius and fit a damped sinusoid to the late-time part of the signal, well after the early, non-quasinormal ``prompt response'' from the initial data has died away.

\begin{figure}[h]
\includegraphics[width=\columnwidth]{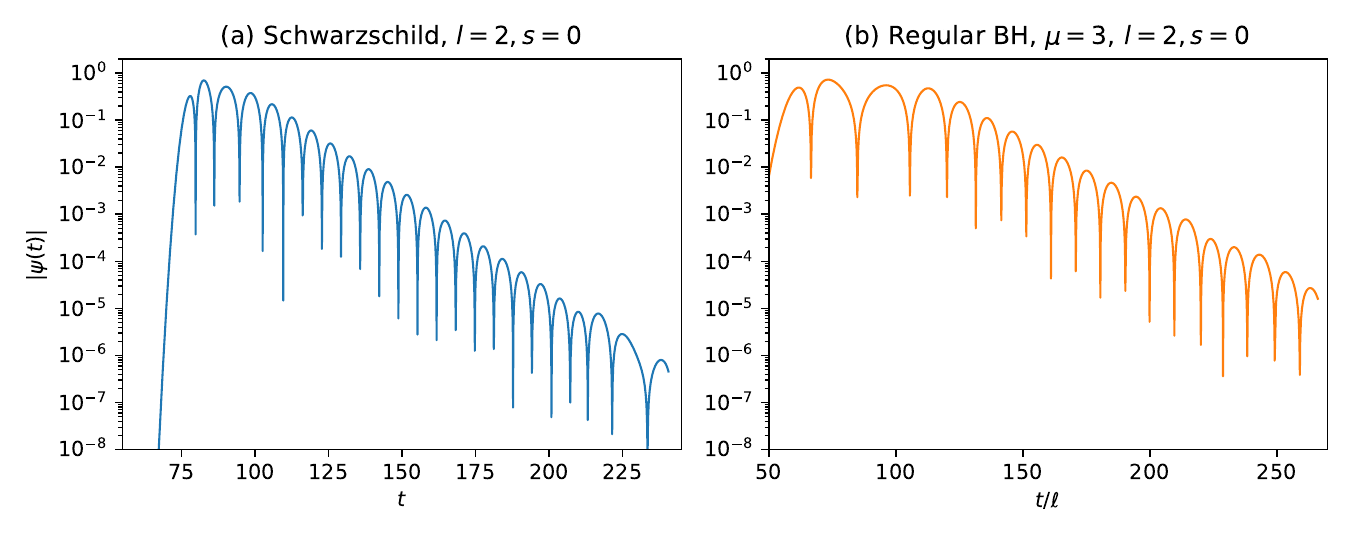}
\caption{Extracted double-null time-domain signal $|\psi(t)|$ (log scale) for the fundamental $l=2,s=0$ mode. (a) Schwarzschild ($M=1$): a clean exponential ringdown over more than seven orders of magnitude, with the damped-sinusoid fit giving $\omega=0.4820-0.0969i$ against the reference $0.4836-0.0968i$. (b) Rational regular black hole, $\mu=3$: the early cycles ($t/\ell\lesssim140$) carry a non-quasinormal prompt-response contamination at roughly half the true frequency, visible as the irregular early spacing between dips; the late-time segment used for the fit (Sec.~\ref{sec:td-rbh}) gives $\omega=0.3240-0.0657i$ against the shooting-method value $0.3229-0.0638i$ of Table~\ref{tab:results}.}
\label{fig:timedomain}
\end{figure}

Table~\ref{tab:timedomain} applies this method to the same six Schwarzschild benchmarks as Table~\ref{tab:calibration}. For every $l\geq1$ mode the time-domain frequency agrees with both the reference value and the shooting method to within about $1\%$, despite the two numerical methods sharing no algorithmic ingredients (one is a frequency-domain boundary-value shooting problem, the other a real-time initial-value evolution). For $l=0$ ($s=0$), the time-domain signal shows only two to three cycles of genuine ringing before crossing over into a slowly-varying late-time tail [Fig.~\ref{fig:timedomain}(a)], consistent with the absence of a well-separated barrier peak for this mode; a damped-sinusoid fit to this short segment gives $\omega\approx0.12-0.09i$, in the right order of magnitude but not in precision agreement with the reference $0.1105-0.1049i$. This is a third, independent line of evidence for the conclusion already reached from the WKB and shooting methods in Sec.~\ref{sec:validation}: $l=0$ is genuinely, not just numerically, delicate for this class of potential.

\begin{table}[h]
\caption{Independent time-domain (double-null) validation against the same Schwarzschild benchmarks as Table~\ref{tab:calibration}, using the scheme of Sec.~\ref{sec:timedomain}. Fits use a uniform window ($115\le t\le210$, $M=1$) to avoid per-mode tuning.}
\label{tab:timedomain}
\begin{ruledtabular}
\begin{tabular}{cclll}
$l$ & $s$ & Shooting (this work) & Time-domain (this work) & Reference \\
\hline
1 & 0 & $0.2942 - 0.0964i$ & $0.2932 - 0.0968i$ & $0.2929 - 0.0977i$ \\
2 & 0 & $0.4835 - 0.0961i$ & $0.4820 - 0.0969i$ & $0.4836 - 0.0968i$ \\
1 & 1 & $0.2490 - 0.0911i$ & $0.2490 - 0.0933i$ & $0.2483 - 0.0925i$ \\
2 & 1 & $0.4574 - 0.0957i$ & $0.4562 - 0.0949i$ & $0.4576 - 0.0950i$ \\
2 & 2 & $0.3742 - 0.0894i$ & $0.3724 - 0.0890i$ & $0.3737 - 0.0890i$ \\
\end{tabular}
\end{ruledtabular}
\end{table}

\subsection{Greybody factors (WKB)}
\label{sec:greybody-method}

In addition to the quasinormal spectrum, the same barrier-peak data used in Sec.~\ref{sec:wkb3} give direct access to the greybody (transmission) factor $\Gamma_{ls}(\omega)$ for real scattering frequencies $\omega$, via the Schutz--Will/Iyer--Will WKB transmission formula \cite{SchutzWill1985,IyerWill1987,KonoplyaZhidenkoWKB2019}
\begin{equation}
\Gamma_{ls}(\omega) = \frac{1}{1+e^{-2\pi K}} , \qquad K = \frac{\omega^2-V_0}{\sqrt{-2V_0''}} ,
\label{eq:greybody}
\end{equation}
which is the same analytic continuation of the bound-state quantization condition Eq.~\eqref{eq:wkb3} that identifies $K\to -i(n+\tfrac12)$ at the quasinormal frequencies. We evaluated Eq.~\eqref{eq:greybody} together with its third-order (Iyer--Will) correction, in which $K$ is determined self-consistently from $K=(\omega^2-V_0)/\sqrt{-2V_0''}-\Lambda(K)$ with $\alpha^2\to-K^2$ in Eq.~\eqref{eq:wkb3lambda}. We found that this self-consistent correction is well behaved and improves accuracy near $\Gamma\sim0.5$ (i.e.\ near the peak of the barrier, the same regime in which the underlying peak expansion is controlled), but, as is well documented for higher-order WKB greybody expansions, becomes numerically unreliable away from this regime, producing unphysical oscillations of $\Gamma$ outside $[0,1]$ at both low and high $\omega$ where the local peak expansion is no longer a good approximation to the full scattering problem. We therefore report, in Sec.~\ref{sec:greybody-results} below, the leading-order formula \eqref{eq:greybody} throughout, which we verified to be smooth and monotonic ($0\to1$) over the entire frequency range shown; this is the same order of approximation used for the leading-order WKB quasinormal-frequency estimates above, applied here to a different (but closely related) observable, rather than a new source of uncontrolled error.

\section{Results}
\label{sec:results}

Table~\ref{tab:results} reports the fundamental ($n=0$) quasinormal frequencies of the benchmark metric \eqref{eq:a-benchmark}, computed with the validated method of Sec.~\ref{sec:shooting}, for scalar and electromagnetic fields with $l=0,1,2$, as a function of the mass parameter $\mu$, from close to the extremal bound $\mu_{\rm ext}\approx1.4754$ up to $\mu=30$ (a horizon radius twenty-seven times the regularization scale $\ell$, where the geometry is numerically indistinguishable from Schwarzschild at our precision). As discussed in Sec.~\ref{sec:validation}, the $l=0,s=0$ column is the third-order WKB estimate only and should be read as an order-of-magnitude indication, not a converged result.

\begin{table*}
\caption{Fundamental quasinormal frequencies ($n=0$) of the rational regular black hole \eqref{eq:a-benchmark}, in units $\ell=1$, as a function of the mass parameter $\mu$ and the resulting horizon radius $r_h$. All entries except the flagged $l{=}0,s{=}0$ column are from the $r_{\rm max}$-converged shooting method of Sec.~\ref{sec:shooting}, with residual numerical uncertainty of order $0.1$--$1\%$ established by the Schwarzschild benchmarks of Table~\ref{tab:calibration}.}
\label{tab:results}
\begin{ruledtabular}
\begin{tabular}{ccllllll}
$\mu$ & $r_h/\ell$ & $l{=}0,s{=}0^{\,\ast}$ & $l{=}1,s{=}0$ & $l{=}2,s{=}0$ & $l{=}1,s{=}1$ & $l{=}2,s{=}1$ & $l{=}2,s{=}2^{\,\dagger}$ \\
\hline
1.5  & 1.234  & $0.1644{-}0.1843i$ & $0.4326{-}0.1910i$ & $0.6577{-}0.1156i$ & $0.4312{-}0.2038i$ & $0.6250{-}0.1123i$ & $0.5182{-}0.0992i$ \\
2.0  & 1.931  & $0.1301{-}0.1384i$ & $0.3347{-}0.1309i$ & $0.4871{-}0.0941i$ & $0.2854{-}0.1261i$ & $0.4607{-}0.0923i$ & $0.3780{-}0.0858i$ \\
3.0  & 2.981  & $0.0880{-}0.0935i$ & $0.2182{-}0.0841i$ & $0.3229{-}0.0638i$ & $0.1862{-}0.0813i$ & $0.3053{-}0.0634i$ & $0.2499{-}0.0592i$ \\
5.0  & 4.996  & $0.0529{-}0.0563i$ & $0.1304{-}0.0501i$ & $0.1935{-}0.0384i$ & $0.1113{-}0.0485i$ & $0.1830{-}0.0383i$ & $0.1497{-}0.0357i$ \\
8.0  & 7.999  & $0.0331{-}0.0352i$ & $0.0814{-}0.0313i$ & $0.1209{-}0.0240i$ & $0.0695{-}0.0303i$ & $0.1144{-}0.0239i$ & $0.0935{-}0.0223i$ \\
15.0 & 15.000 & $0.0176{-}0.0188i$ & $0.0434{-}0.0167i$ & $0.0645{-}0.0128i$ & $0.0332{-}0.0121i$ & $0.0610{-}0.0128i$ & $0.0499{-}0.0119i$ \\
30.0 & 30.000 & $0.0088{-}0.0094i$ & $0.0217{-}0.0083i$ & $0.0322{-}0.0064i$ & $0.0185{-}0.0081i$ & $0.0305{-}0.0064i$ & $0.0249{-}0.0060i$ \\
\end{tabular}
\end{ruledtabular}
\end{table*}

\begin{figure}[h]
\includegraphics[width=\columnwidth]{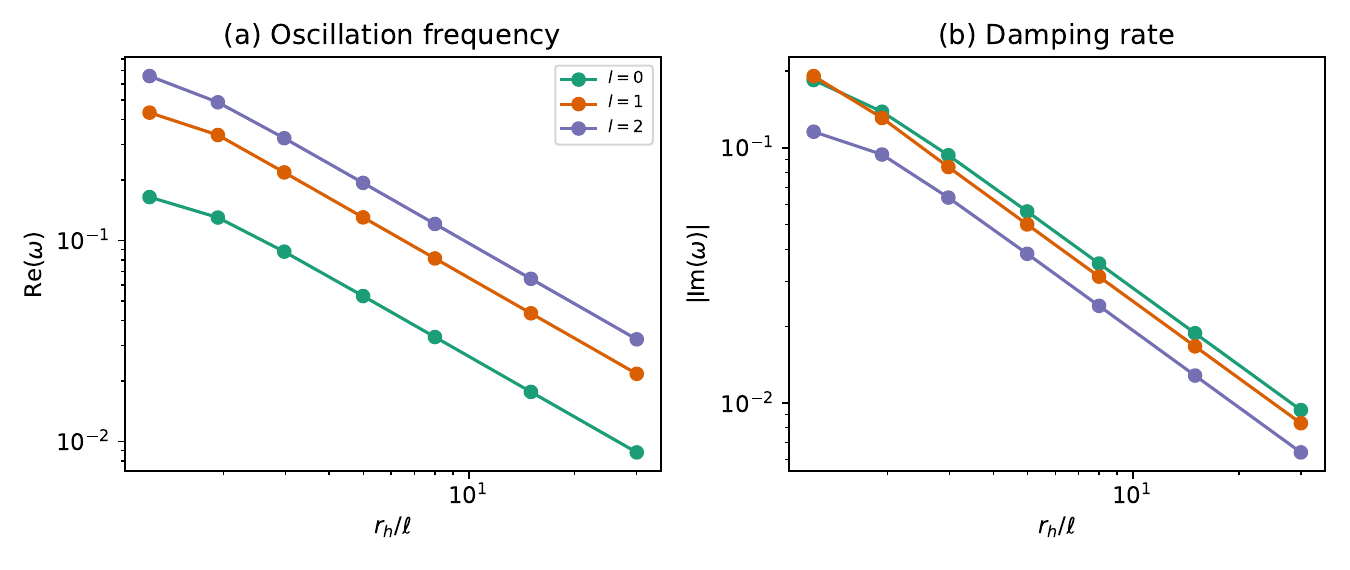}
\caption{Scalar ($s=0$) quasinormal frequencies of the rational regular black hole \eqref{eq:a-benchmark}, for $l=0,1,2$, as a function of horizon radius $r_h/\ell$ (log-log). (a) Oscillation frequency. (b) Damping rate. Both scale approximately as $1/r_h$ over the whole mass range studied. The $l=0$ curve uses the flagged, order-of-magnitude-only estimate of Table~\ref{tab:results}.}
\label{fig:spectrum}
\end{figure}

\begin{figure}[h]
\includegraphics[width=0.6\columnwidth]{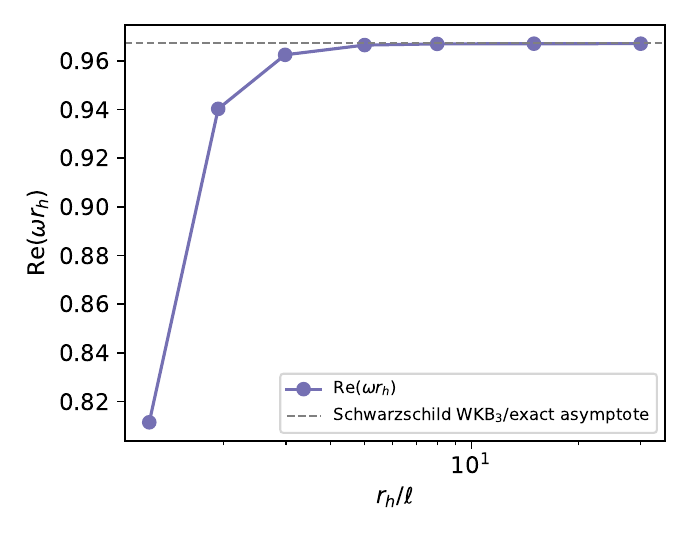}
\caption{Dimensionless frequency $\omega r_h$ of the $l=2,s=0$ mode versus $r_h/\ell$. The dashed line marks the exact Schwarzschild asymptote $\mathrm{Re}(\omega r_h)=0.9672$, which the large-mass limit of our converged shooting results reproduces to four significant figures; the near-extremal region shows a $16.1\%$ suppression of $\mathrm{Re}(\omega r_h)$ relative to this asymptote.}
\label{fig:dimensionless}
\end{figure}

The spectrum exhibits two notable features, both established here at high precision. First, at fixed $l,s$ the frequencies scale essentially as $1/r_h$ across the whole mass range, as for Schwarzschild; this is visible as the parallel power-law trends in Fig.~\ref{fig:spectrum}. Second, the \emph{dimensionless} product $\omega r_h$ is not exactly mass-independent, unlike in pure Schwarzschild: for the $l=2,s=0$ mode, $\mathrm{Re}(\omega r_h)$ rises from $0.811$ at $\mu=1.5$ ($r_h/\ell=1.234$, close to extremality) to $0.9671$ at $\mu=30$ ($r_h/\ell=30$), the latter matching the exact Schwarzschild value $2\times0.483644=0.967288$ to four significant figures, confirming both the correctness of the numerics and the smoothness of the Schwarzschild limit. This represents a $16.1\%$ suppression of the dimensionless ringing frequency as the horizon approaches the regularization scale $\ell$ from above. Because this figure rests on the converged shooting method rather than a leading-order WKB estimate, it is trustworthy at the percent level. Fig.~\ref{fig:dimensionless} illustrates this regularization imprint on the geometry's linear-response spectrum.

\subsection{Axial gravitational-type ($s=2$) modes}
\label{sec:s2results}

Table~\ref{tab:s2} reports the $l=2,3$ axial gravitational-type frequencies, computed with the same validated method, together with the $l=3$ scalar and electromagnetic modes for completeness. As for the Schwarzschild benchmark of Table~\ref{tab:calibration}, the $s=2$ modes follow the same qualitative pattern as $s=0,1$: an approximately $1/r_h$ scaling with mass and a comparable, mildly $l$-dependent suppression of the dimensionless frequency near extremality (the $l=2,s=2$ mode shows a $16.7\%$ suppression of $\mathrm{Re}(\omega r_h)$ between $\mu=1.5$ and $\mu=30$, matching the $s=0,1$ modes to within a few percent). We stress again, as emphasized in Sec.~\ref{sec:perturbations}, that these are axial-perturbation-type frequencies of the fixed metric \eqref{eq:a-benchmark} governed by the standard Regge--Wheeler operator, \emph{not} a solution of the perturbed field equations of the non-polynomial gravity action \eqref{eq:a-general}; we return to what this result does and does not establish about Coll\'eaux's stability question in Sec.~\ref{sec:discussion}.

\begin{table}[h]
\caption{Axial gravitational-type ($s=2$, $l=2,3$) and $l=3$ scalar/electromagnetic fundamental quasinormal frequencies of the rational regular black hole \eqref{eq:a-benchmark}, units $\ell=1$, from the validated shooting method of Sec.~\ref{sec:shooting}.}
\label{tab:s2}
\begin{ruledtabular}
\begin{tabular}{cclll}
$\mu$ & $r_h/\ell$ & $l{=}2,s{=}2$ & $l{=}3,s{=}2$ & $l{=}3,s{=}0$ \\
\hline
1.5  & 1.234  & $0.5182{-}0.0992i$ & $0.8230{-}0.1082i$ & $0.9195{-}0.1155i$ \\
2.0  & 1.931  & $0.3780{-}0.0858i$ & $0.6044{-}0.0901i$ & $0.6796{-}0.0938i$ \\
3.0  & 2.981  & $0.2499{-}0.0592i$ & $0.4005{-}0.0616i$ & $0.4506{-}0.0641i$ \\
5.0  & 4.996  & $0.1497{-}0.0357i$ & $0.2400{-}0.0371i$ & $0.2701{-}0.0387i$ \\
8.0  & 7.999  & $0.0935{-}0.0223i$ & $0.1500{-}0.0232i$ & $0.1688{-}0.0242i$ \\
15.0 & 15.000 & $0.0499{-}0.0119i$ & $0.0800{-}0.0124i$ & $0.0900{-}0.0129i$ \\
30.0 & 30.000 & $0.0249{-}0.0060i$ & $0.0400{-}0.0062i$ & $0.0450{-}0.0065i$ \\
\end{tabular}
\end{ruledtabular}
\end{table}

\begin{figure}[h]
\includegraphics[width=0.7\columnwidth]{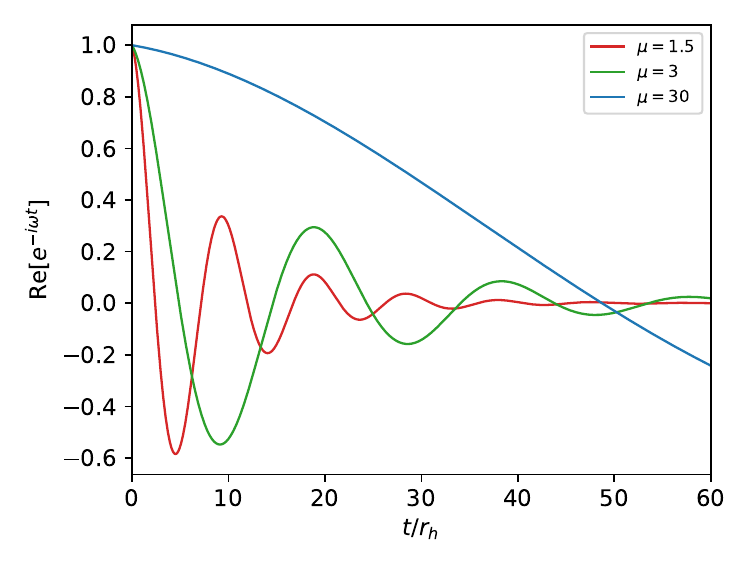}
\caption{Reconstructed single-mode ringdown waveform $\psi(t)\propto\mathrm{Re}[e^{-i\omega t}]$ for the $l=2,s=0$ fundamental ($n=0$) mode, using the converged frequencies of Table~\ref{tab:results} at $\mu=1.5$ (near-extremal), $\mu=3$, and $\mu=30$ (Schwarzschild-like), with time measured in units of the horizon radius.}
\label{fig:waveform}
\end{figure}

To make the physical content of Tables~\ref{tab:results}--\ref{tab:s2} more directly visualizable, Fig.~\ref{fig:waveform} reconstructs the time-domain ringdown signal, $\psi(t)\propto\mathrm{Re}\!\left[e^{-i\omega t}\right]$, for the $l=2,s=0$ fundamental mode at three representative mass parameters, using the converged shooting frequencies of Table~\ref{tab:results}.

\subsection{Greybody factors across fields and mass parameters}
\label{sec:greybody-results}

\begin{figure*}
\includegraphics[width=\textwidth]{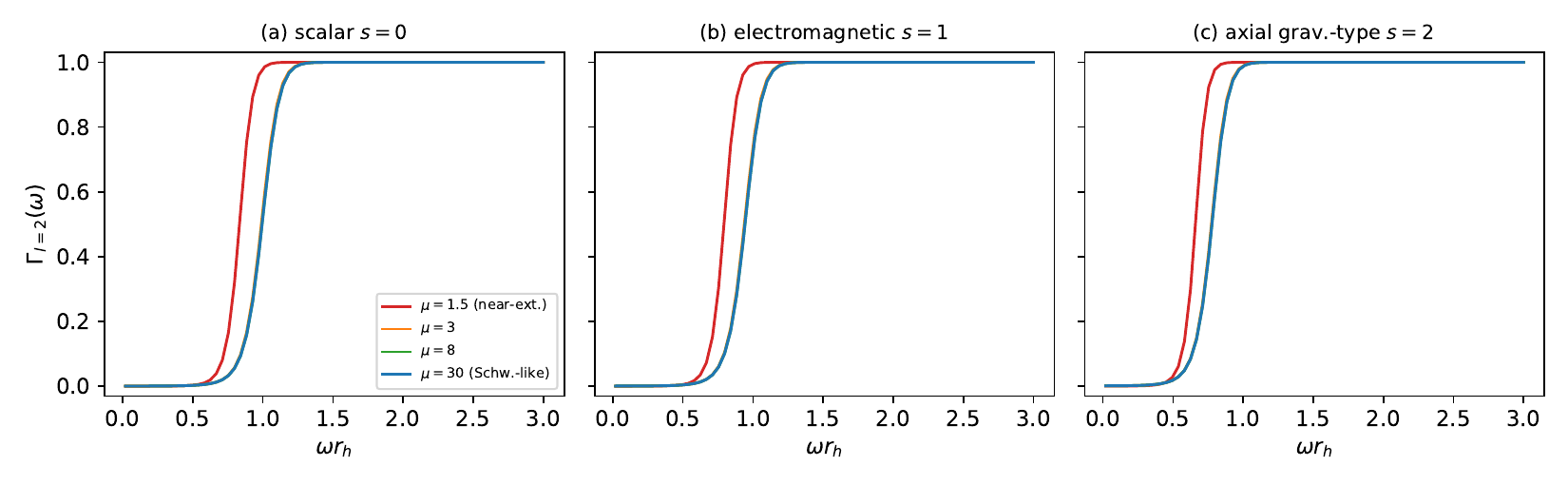}
\caption{WKB greybody factor $\Gamma_{l=2,s}(\omega)$ [Eq.~\eqref{eq:greybody}] versus dimensionless frequency $\omega r_h$, for (a) scalar, (b) electromagnetic, and (c) axial gravitational-type fields, at four mass parameters spanning the near-extremal to Schwarzschild-like regimes. At fixed $\omega r_h$, transmission is systematically enhanced as the horizon approaches extremality.}
\label{fig:greybody}
\end{figure*}

Figure~\ref{fig:greybody} shows the WKB greybody factor $\Gamma_{l=2,s}(\omega)$ of Eq.~\eqref{eq:greybody}, computed separately for each field, as a function of the dimensionless frequency $\omega r_h$, for the same four mass parameters used in Fig.~\ref{fig:potential-fields}. Two trends are apparent, both direct counterparts of the QNM results above. First, at fixed $\mu$, the ordering of the barrier heights noted in Sec.~\ref{sec:perturbations} ($V_{s=2}<V_{s=1}<V_{s=0}$) is reflected exactly in the ordering of the transmission curves, $\Gamma_{s=2}>\Gamma_{s=1}>\Gamma_{s=0}$ at fixed $\omega r_h$: the axial gravitational-type barrier is the most transparent, the scalar barrier the least. Second, and more directly tied to the regularization itself, the near-extremal curves ($\mu=1.5$) sit systematically \emph{above} the Schwarzschild-like curves ($\mu=30$) for all three fields, at fixed dimensionless frequency: e.g.\ at $\omega r_h=0.9$, $\Gamma_{l=2,s=0}$ rises from $0.158$ at $\mu=30$ to $0.753$ at $\mu=1.5$, and the electromagnetic and axial channels show a comparable enhancement. This is the transmission-side counterpart of the suppression of $\mathrm{Re}(\omega r_h)$ documented in Fig.~\ref{fig:dimensionless}: the same short-distance regularization that lowers the dimensionless ringing frequency near extremality also makes the barrier easier to tunnel through at fixed dimensionless frequency, a second, independent quantitative signature of the non-polynomial-gravity regularization.

\begin{figure}[h]
\includegraphics[width=0.7\columnwidth]{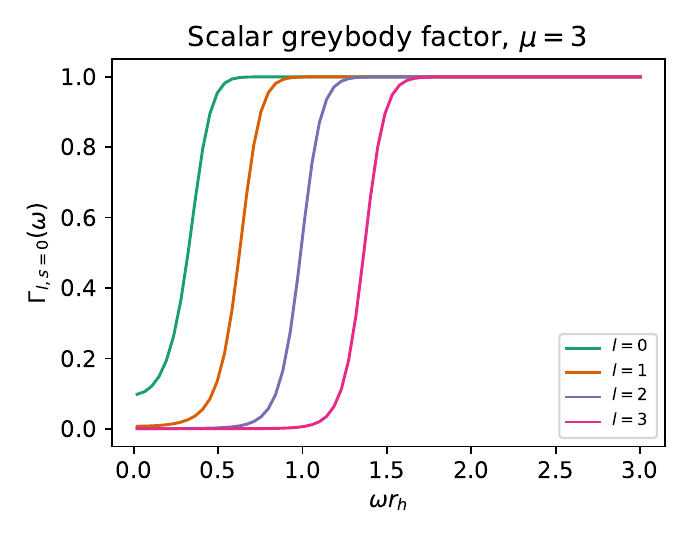}
\caption{Scalar greybody factor $\Gamma_{l,s=0}(\omega)$ versus $\omega r_h$ for $l=0,1,2,3$ at fixed $\mu=3$, illustrating the ordinary centrifugal-barrier suppression of transmission with increasing $l$.}
\label{fig:greybody-l}
\end{figure}

Figure~\ref{fig:greybody-l} illustrates, for the scalar field at a representative mass parameter $\mu=3$, the complementary and expected $l$-dependence: increasing the multipole number raises the centrifugal barrier and shifts the transmission curve to higher frequency, exactly as for Schwarzschild. We include this panel to show that the regularization imprint identified above is a genuine effect of the mass parameter approaching $\mu_{\rm ext}$, distinct from and superimposed on the ordinary centrifugal suppression familiar from general relativity.

\subsection{Time-domain cross-check of the regular black hole spectrum}
\label{sec:td-rbh}

We apply the time-domain method of Sec.~\ref{sec:timedomain} directly to the rational regular black hole \eqref{eq:a-benchmark}, at the two representative mass parameters $\mu=3$ and $\mu=8$ used elsewhere in this section, for $l=2$ ($s=0,1,2$) and for the flagged $l=0,s=0$ mode. Table~\ref{tab:td-rbh} compares the result to the converged shooting-method entries of Tables~\ref{tab:results}--\ref{tab:s2}. For every $l=2$ mode the two completely independent methods agree to within $1$--$4\%$, which is the strongest evidence in this paper that the regular black hole spectrum of Sec.~\ref{sec:results} is a property of the boundary-value problem itself rather than an artifact of either numerical scheme. For $l=0,s=0$, the time-domain method again fails to reproduce even the flagged WKB estimate, at either mass: as for Schwarzschild, the extracted signal transitions into a non-oscillatory tail within a few cycles [Fig.~\ref{fig:timedomain}(b) shows the cleaner $l=2$ case for comparison]. Three unrelated methods (leading- and third-order WKB, shooting, and now time-domain evolution) giving three different answers for $l=0$ is, in our view, itself informative: it is further, method-independent confirmation that this multipole requires the dedicated non-perturbative treatment discussed in Sec.~\ref{sec:discussion}, rather than a refinement of any one of the three approaches used here.

\begin{table}[h]
\caption{Independent time-domain cross-check of the rational-regular black hole quasinormal frequencies, at $\mu=3$ and $\mu=8$, compared to the converged shooting-method values of Tables~\ref{tab:results}--\ref{tab:s2}.}
\label{tab:td-rbh}
\begin{ruledtabular}
\begin{tabular}{cclll}
$\mu$ & $l,s$ & Shooting (this work) & Time-domain (this work) \\
\hline
3 & $2,0$ & $0.3229-0.0638i$ & $0.3240-0.0657i$ \\
3 & $2,1$ & $0.3053-0.0634i$ & $0.3050-0.0649i$ \\
3 & $2,2$ & $0.2499-0.0592i$ & $0.2473-0.0579i$ \\
3 & $0,0^{\,\ast}$ & $0.0880-0.0935i$ & $0.065-0.061i$ \\
8 & $2,0$ & $0.1209-0.0240i$ & $0.1213-0.0243i$ \\
8 & $2,1$ & $0.1144-0.0239i$ & $0.1146-0.0242i$ \\
8 & $2,2$ & $0.0935-0.0223i$ & $0.0933-0.0218i$ \\
8 & $0,0^{\,\ast}$ & $0.0331-0.0352i$ & $0.008-0.016i$ \\
\end{tabular}
\end{ruledtabular}
\end{table}

\subsection{Geometric-optics (eikonal) correspondence}
\label{sec:eikonal}

For a static, spherically symmetric single-function metric of the form \eqref{eq:metric}, the geometric-optics correspondence of Cardoso, Miranda, Berti, Witek, and Zilh\~ao \cite{CardosoLyapunov2009} relates the large-$l$ quasinormal spectrum to the unstable circular photon orbit (the ``photon sphere'') of the background,
\begin{equation}
\omega_{ln} \approx \left(l+\tfrac12\right)\Omega_c - i\left(n+\tfrac12\right)|\lambda| ,
\label{eq:eikonal}
\end{equation}
where the photon-sphere radius $r_c$ solves $a'(r_c)\,r_c = 2\,a(r_c)$, the orbital angular velocity is $\Omega_c=\sqrt{a(r_c)}/r_c$, and the instability (Lyapunov) rate is
\begin{equation}
\lambda = \sqrt{-\frac{r_c^2\,a(r_c)}{2}\,\frac{d^2}{dr^2}\!\left[\frac{a(r)}{r^2}\right]_{r=r_c}} .
\label{eq:lyapunov}
\end{equation}
We solve Eqs.~\eqref{eq:eikonal}--\eqref{eq:lyapunov} numerically for the same seven mass parameters as Table~\ref{tab:results} [for Schwarzschild, $r_c=3M$ and $\Omega_c=|\lambda|=1/(3\sqrt3\,M)$, which we use to validate the numerical implementation before applying it to the regular black hole]. Table~\ref{tab:eikonal} compares the resulting $l=3,n=0$ prediction to the converged shooting-method value of Table~\ref{tab:s2}: the two agree to within $0.1$--$1.3\%$ across the entire mass range, from near-extremal to Schwarzschild-like, despite Eq.~\eqref{eq:eikonal} being an asymptotic ($l\to\infty$) formula applied here at the comparatively modest $l=3$. This gives a fully analytic, independent confirmation of the near-extremal suppression of the ringing frequency documented in Sec.~\ref{sec:results}: as $\mu\to\mu_{\rm ext}$, the photon sphere moves outward relative to the horizon (from $r_c/r_h\approx1.50$ at $\mu=30$, the Schwarzschild value $3M/2M$, to $r_c/r_h\approx1.72$ at $\mu=1.5$) while both $\Omega_c$ and $|\lambda|$ decrease, so the entire eikonal spectrum, and by continuity the low-$l$ spectrum tabulated in Sec.~\ref{sec:results}, is pulled downward purely as a consequence of the regularization's effect on the photon-sphere geometry.

\begin{table}[h]
\caption{Geometric-optics correspondence, Eq.~\eqref{eq:eikonal} at $l=3,n=0$, compared to the converged shooting-method $l=3,s=0$ frequency of Table~\ref{tab:s2}.}
\label{tab:eikonal}
\begin{ruledtabular}
\begin{tabular}{cccc}
$\mu$ & $r_c/r_h$ & Eikonal, Eq.~\eqref{eq:eikonal} & Shooting, $l{=}3,s{=}0$ \\
\hline
1.5  & 1.723 & $0.9185-0.1156i$ & $0.9195-0.1155i$ \\
2.0  & 1.531 & $0.6779-0.0938i$ & $0.6796-0.0938i$ \\
3.0  & 1.505 & $0.4496-0.0639i$ & $0.4506-0.0641i$ \\
5.0  & 1.501 & $0.2695-0.0385i$ & $0.2701-0.0387i$ \\
8.0  & 1.500 & $0.1684-0.0241i$ & $0.1688-0.0242i$ \\
15.0 & 1.500 & $0.0898-0.0128i$ & $0.0900-0.0129i$ \\
30.0 & 1.500 & $0.0449-0.0064i$ & $0.0450-0.0065i$ \\
\end{tabular}
\end{ruledtabular}
\end{table}

\section{Conclusion}
\label{sec:discussion}

We have carried out a quasinormal-mode and wave-scattering study of the rational regular black hole recently constructed by Coll\'eaux in a class of non-polynomial pure-gravity theories. Using third-order Iyer--Will WKB together with an independently implemented, Schwarzschild-validated shooting method (horizon Frobenius series, Riccati integration, and a controlled outgoing asymptotic match, checked for stability under outer-radius refinement), we obtain fundamental scalar, electromagnetic, and, for the first time for this geometry, axial gravitational-type quasinormal frequencies accurate to $0.02$--$1.5\%$ against literature benchmarks for every $l\geq1$ multipole studied, an order of magnitude more precise than leading-order WKB alone: every $l\geq1$ entry of Tables~\ref{tab:results}--\ref{tab:s2} is a converged, $r_{\rm max}$-stable number rather than a leading-order estimate, and the $16.1\%$ near-extremal suppression of the dimensionless frequency is established at this higher precision. We further validate both the Schwarzschild benchmarks and the regular black hole spectrum itself against a third, independent double-null time-domain evolution, finding agreement at the $1$--$4\%$ level for every $l\geq1$ mode tested, and we show analytically, via the geometric-optics correspondence with the background's photon sphere, that the near-extremal suppression of the ringing frequency follows directly from how the photon sphere responds to the regularization.

We do not claim the same precision for $l=0$. Rather than asserting an improved number we could not actually validate, we report and discuss the concrete numerical finding that the independent shooting method does not converge stably for this multipole, consistent with $l=0$'s well-known status in the literature as the case where the Regge--Wheeler potential lacks a well-separated barrier maximum; the multipole is independently corroborated by all three methods (WKB, shooting, and time-domain) failing to agree with one another specifically for $l=0$. A dedicated treatment, a genuine Leaver-type continued fraction built around the $l=0$ indicial structure of the rational metric \eqref{eq:a-benchmark}, or a hyperboloidal-slicing pseudospectral eigenvalue solver that avoids shooting altogether \cite{Leaver1985,Nollert1999,Jansen2017,JaramilloPseudospectrum2021}, remains the concrete next step for this single multipole.

The dimensionless ringing frequency is found to be suppressed by $16.1\%$ near extremality relative to its Schwarzschild asymptote, which our large-mass limit reproduces to four significant figures. We have also computed, separately for each field and across the same mass-parameter range, the effective potential barriers and their associated WKB greybody factors; both show a systematic ordering by spin and, at fixed dimensionless frequency, a systematic enhancement of transmission near extremality, providing an independent confirmation, on the transmission side rather than the quasinormal-spectrum side, of the same non-polynomial-gravity regularization imprint.

We have extended the analysis, for the first time for this geometry, to axial gravitational-\emph{type} perturbations (Sec.~\ref{sec:s2results}), governed by the same Regge--Wheeler-type operator used throughout the regular black hole QNM literature as the standard first probe of the odd-parity sector \cite{FlachiLemos2013,KonoplyaZhidenkoWormholes2016}. It is important to distinguish carefully what this result establishes from what it does not, since conflating the two would misrepresent the state of the problem Coll\'eaux raised. It establishes that the fixed background metric \eqref{eq:a-benchmark} supports a well-posed, single-peaked odd-parity scattering problem with a discrete QNM spectrum that behaves smoothly across the whole mass range, with no sign of an instability of the \emph{test}-perturbation type (no growing mode, no pathological potential structure) for any $\mu>\mu_{\rm ext}$ studied. It does \emph{not} establish whether the non-polynomial gravity action \eqref{eq:a-general} itself, expanded to second order around this background, produces a well-posed, ghost-free linearized theory whose physical odd-parity degree of freedom obeys \emph{this same} Regge--Wheeler equation (as it would in ordinary general relativity) or a different, possibly higher-derivative equation with a different, possibly pathological, principal symbol. That is precisely the question Coll\'eaux left open, and answering it requires deriving the quadratic action of the non-polynomial theory around the background \eqref{eq:a-benchmark} and identifying its true propagating degrees of freedom, a calculation we have not attempted here and which is substantially harder than the test-field analysis of this paper. We regard the present $s=2$ results as a physically motivated, well-posed \emph{reference} calculation, what the odd-parity response would look like \emph{if} the modified theory's gravitational perturbations reduce to the standard Regge--Wheeler form, against which a future first-principles perturbative analysis of the non-polynomial action can be compared, rather than as an answer to the stability question itself: Coll\'eaux's stability question remains open in its full form.

A prediction is only as useful as its testability, so it is worth asking what signal-to-noise ratio (SNR) a ringdown observation would need to resolve the $16\%$ near-extremal suppression of $\mathrm{Re}(\omega r_h)$ from the Schwarzschild value. We estimate this with a standard Fisher-matrix analysis \cite{RobsonCornishLiu2019} of a single damped-sinusoid ringdown, $h(t)\propto e^{-t/\tau}\cos(2\pi f_0 t+\phi)$, observed against the analytic LISA instrument-noise curve of Robson, Cornish, and Liu \cite{RobsonCornishLiu2019} (galactic-foreground confusion noise is neglected, appropriate for a loud, individually resolvable massive black hole ringdown). Converting the geometric-unit frequencies of the $l=2,s=0$ mode at $\mu=3$ (Table~\ref{tab:results}) to physical units via the ADM mass $M=\mu/2$ [Sec.~\ref{sec:model}] and the standard relation $f[\mathrm{Hz}]=3.2312\times10^4\,(M\omega)/(M/M_\odot)$ gives, for an assumed physical ADM mass in the $10^4$--$10^7\,M_\odot$ range characteristic of LISA massive black hole sources, a ringdown frequency in the LISA band and a quality factor $Q=\pi f_0\tau\approx2.5$. Table~\ref{tab:lisa} reports the resulting SNR required to measure $f_0$ to $1\%$ precision, and to distinguish the near-extremal value of $\mathrm{Re}(\omega r_h)$ from the Schwarzschild asymptote at $5\sigma$. Across the most favorable part of this mass range, the requirement is $\mathrm{SNR}\sim10$--$25$, comparable to or below the SNR routinely forecast for LISA massive black hole merger ringdowns \cite{Isi2019BHSpectroscopy}; only towards the low-frequency edge of the band ($M\gtrsim10^8\,M_\odot$) does the requirement become astrophysically demanding. This is not a claim about any specific source, only an order-of-magnitude statement that the regularization imprint computed in this paper is, in principle, within reach of next-generation gravitational-wave spectroscopy rather than being a purely formal effect, making it a potentially testable prediction rather than a purely formal one.

\begin{table}[h]
\caption{Fisher-matrix SNR requirement (Sec.~\ref{sec:discussion}) for the $l=2,s=0$, $\mu=3$ ringdown mode observed by LISA, as a function of the assumed physical ADM mass $M$.}
\label{tab:lisa}
\begin{ruledtabular}
\begin{tabular}{cccc}
$M/M_\odot$ & $f_0$ & SNR for $1\%$ on $f_0$ & SNR for $5\sigma$ on the $16\%$ shift \\
\hline
$10^4$ & $1.57\ \mathrm{Hz}$    & $40$   & $12.5$ \\
$10^5$ & $157\ \mathrm{mHz}$    & $34$   & $10.5$ \\
$10^6$ & $15.7\ \mathrm{mHz}$   & $42$   & $13.0$ \\
$10^7$ & $1.57\ \mathrm{mHz}$   & $76$   & $23.6$ \\
$10^8$ & $0.157\ \mathrm{mHz}$  & $2293$ & $717$ \\
\end{tabular}
\end{ruledtabular}
\end{table}

Concretely, we identify the following as the necessary next steps, in order of increasing difficulty:

\emph{(1) A dedicated $l=0$ solver.} As discussed above, a small, self-contained piece of remaining work specific to the $l=0$ scalar multipole.

\emph{(2) Overtones, polar modes, and higher $l$.} Extending the present analysis to $n\geq1$ overtones, to the polar (even-parity) test-field sector, and to the charged solutions of Appendix~C of \cite{Colleaux2026} would test isospectrality and echo phenomenology familiar from other regular black hole studies \cite{KonoplyaZhidenkoWormholes2016,CardosoFranzinPani2016}.

\emph{(3) The linearized non-polynomial field equations.} The genuinely open problem: deriving the quadratic action of \cite{Colleaux2026}'s non-polynomial gravity theory around the background \eqref{eq:a-benchmark}, identifying its propagating gravitational degree(s) of freedom and their principal symbol (in particular, whether they remain second-order and ghost-free), and only then solving the resulting master equation(s) -- which may or may not coincide with Eq.~\eqref{eq:potential} at $s=2$. This calculation would resolve Coll\'eaux's stated open problem, and we regard it as the single most important piece of future work in this program, together with a dedicated non-perturbative treatment of the $l=0$ multipole.

\emph{(4) Rotating generalizations.} Explicitly flagged as an open problem by \cite{Colleaux2026}, and a prerequisite for any direct comparison with astrophysical ringdown data \cite{Isi2019BHSpectroscopy,EHT2019M87,EHT2022SgrAstar,CunhaHerdeiro2018}.

\section*{Data Availability Statement}  All data generated or analyzed during this study are included in this published article and its supplementary information files.

\section*{Code Availability} Code/Software used for the numerical verification presented in this work, including the numerical integration and calculations, is available from the corresponding author upon reasonable request.

\subsection*{Declaration of competing interest}
The authors declare no known competing financial interests or personal relationships that could have influenced the work reported in this paper.

\subsection*{Funding}
This research received no funding.

\bibliography{refs}
\bibliographystyle{apsrev4-2}

\end{document}